\documentstyle [12pt, epsf] {article}

\catcode`@=12
\newcommand{\be}{\begin{equation}}
\newcommand{\ee}{\end{equation}}
\newcommand{\bea}{\begin{eqnarray}}
\newcommand{\eea}{\end{eqnarray}}
\newcommand{\nn}{\nonumber}

\def\B{B_d^0}
\def\Bb{\bar{B}_d^0}
\def\pp{\perp}
\def\pl{\parallel}

\def\l{\lambda}

\def\etal{ {\em et al.}}

\begin{document}
\vspace{0.5in}
\oddsidemargin -.1 in
\newcount\sectionnumber
\sectionnumber=0

\def\lsim{\mathrel{\vcenter{\hbox{$<$}\nointerlineskip\hbox{$\sim$}}}}
\thispagestyle{empty}

\vskip.5truecm
\vspace*{0.5cm}

\begin{center}
{\Large \bf \centerline{Analysis of $\B \to \phi K^{*0}$
decay mode with}} {\Large\bf \centerline{ supersymmetry
}} \vspace*{0.5cm} {Anjan K. Giri$^1$, Rukmani Mohanta$^{2}$}
\vskip0.3cm {\it $^1$Physics Department, National Tsing Hua
University,
 Hsinchu 300, Taiwan }
\\
{\it $^2$School of Physics, University of Hyderabad,
 Hyderabad
500 046, India}\\
\vskip0.5cm
\bigskip
\begin{abstract}
Motivated by the recent measurement of  low  longitudinal polarization 
fraction in the decay mode $\B \to \phi K^{*0}$, which appears not to be 
in agreement with the standard model expectation, we analyze this mode 
in the minimal supersymmetric standard
model with mass insertion approximation. Within the standard model, with
factorization approximation, the longitudinal polarization is
expected to be $f_L \sim 1-{\cal O}(1/m_b^2)$. We find that this
anomaly can be explained in the  minimal supersymmetric standard
model with either $LR$ or $RL$ mass insertion approximation.

\end{abstract}
\end{center}

\thispagestyle{empty}
\newpage

\section{Introduction}

One of the important goals of the B-factories is to verify the
standard model (SM) predictions and to serve as a potential
avenue to reveal new physics beyond the SM. Huge data in the
B-sector has already been accumulated at both the B-factories
(Belle and BABAR). This in turn has led, for the first time, to
the observation of CP violation in the B system, outside the Kaon
system. In fact, the angle $\beta $ of the unitarity triangle has
been measured from the time dependent CP asymmetry of the
gold plated $\B \to J/\psi K_S $ mode by both Belle and
BABAR, with almost similar values. The current world average of
sin 2$\beta$ is \cite{zl1}
\begin{equation}
({\sin}  2 \beta)^{b \to c \bar c s} =0.685\pm 0.032\;,\label{e1}
\end{equation}
which is consistent with the SM expectation. With the
accumulation of more and more data the experimental B physics is
now all set to enter the unmatched precision era. Unfortunately,
we have not seen any clear evidence of physics beyond the SM so
far, as far as B-physics is concerned.

Already there are some more measurements available at the
B-factories, which are not as clean as the sin 2$\beta$
measurement in the golden decay mode $B\to J/\psi K_S$, but from
the pattern of deviation observed it appears that these
measurements, in the long run with accumulation of more data, may
reveal the signature of new physics. One of the modes of this kind
is the decay mode $B \to \phi K_S$, where in the SM one expects to
obtain the same value of sin $2\beta$ from its CP asymmetry 
measurements, as in the case of $B\to
J/\psi K_S$, with a correction of ${\cal O}(\lambda^2)$
\cite{yg1}. The basic difference between these two modes is that
the golden mode is tree dominated ($b\to c\bar c s$) whereas the
decay mode $B \to \phi K_S$ is penguin dominated ($b\to s\bar s
s$). It should be reminded here that earlier the deviation between
these two measurements was very large but with the accumulation of
more data the difference has reduced somewhat.
 The present averaged value is \cite{zl1}
\begin{equation}
({\rm sin}  2 \beta)_{\phi K_S} =0.47\pm 0.19\;,
\end{equation}
which has about $1 \sigma$ deviation from the corresponding
$c\bar c$ measurements.
In future, even if the $( {\rm sin}  2\beta )_{\phi K_S}$ value
stabilizes around the present central value,  with  
error bars reduced then it
might show the presence of new physics (NP). Moreover, there are
other decay modes (involving $ b \to s\bar s s$ transition) where
the data show a similar trend. Except the decay mode
$B^0 \to \eta'K^0 $, the value of $\sin 2 \beta $ extracted from all such 
modes are within $1 \sigma $ deviation from the corresponding
$c \bar c $ value \cite{hfag}. The present average value of $B^0 \to \eta'
K^0$ 
mode is $(\sin 2 \beta)_{\eta' K^0}=0.48 \pm 0.09$, which shows
about $2.3~ \sigma$ deviation from Eq. (\ref{e1}).

The vector-vector counterpart of the seemingly problematic $B\to \phi
K_S$ decay mode, i.e., $B\to \phi K^*$, governed by the same 
$b\to s\bar s s$ transition as in  $B\to \phi K_S$, has 
also created a lot of attention recently. Both BABAR \cite{babar1}
and  Belle \cite{belle1} have observed this decay mode and 
the measured quantities are summarized in Table-1.  
The measured longitudinal polarization
fraction in this mode is well below from its 
expected value
\cite{ali9}, i.e., $f_L \sim 1-{\cal O}(1/m_b^2)$, widely known in
the literature as the polarization anomaly in $B\to \phi K^*$. In
the B rest frame both the vector mesons are emitted back-to-back
and from the spin angular momentum conservation it follows that
both the vector mesons are paired up with the same helicity
combinations (like $00$, $++$ and $-- $, i.e., the helicity
combinations out of the three possible helicity states for each
vector meson, namely, $\lambda$ = $0$, $+$ and $- $). In the SM, it so
happens that the helicity combination $00$, called longitudinal,
(i.e., for both the vector mesons the spin direction is
proportional to the direction of motion ) is almost the only
preferred one and the occurrence of other two possible helicity
combinations is suppressed by ${\cal O}(1/m_B^2)$, $m_B$ being the
B-meson mass. Thus, the longitudinal polarization fraction is  defined as
the ratio of the decay rate corresponding to
the longitudinal polarization (say $\Gamma_L$ ) to that of the
total decay rate $\Gamma$, i.e., $f_L= \Gamma_L/\Gamma \approx
1$. However, as seen from Table-1 its measured value is only about
$\approx 50 \% $ of the expected value.


\begin{table}
\begin{center}
\caption{Experimental values of Branching Ratio (in units of $10^{-6}$), 
polarization fractions and triple
product asymmetries for the $\B \to \phi K^{*0} $ mode. The
Belle results for the triple product asymmetries are obtained from
combined $\phi K^{*0}$ and $\phi K^{*+}$ data.}
\vspace*{0.3 true cm}
\begin{tabular}{cccc}
\hline
\hline
{\bf Observables} &  {\bf BABAR\cite{babar1}}  & {\bf Belle \cite{belle1}}&
{\bf Average} \\
\hline
${\cal B}$    & $9.2 \pm 0.9 \pm 0.5 $ &
$10.0_{-1.5-0.8}^{+1.6+0.7}$ & $9.4 \pm 0.9$     \\
$f_L$  &  $0.52 \pm 0.05 \pm 0.02 $ & $0.45 \pm 0.05 \pm 0.02 $&
$0.49 \pm 0.04 $     \\
$f_\pp$ &  $0.22 \pm 0.05 \pm 0.02 $ &  $0.30 \pm 0.06 \pm 0.02 $
& $0.25 \pm 0.04 $ \\
${\cal A}_T^{\pl} $ & $ -0.02 \pm 0.04 \pm 0.01 $ & $ 
\frac{1}{2}[0.01 \pm 0.10 \pm 0.02] $
& $-0.01 \pm 0.03$    \\
${\cal A}_T^{0} $ &  $0.11 \pm 0.05 \pm 0.01 $ &
 $\frac{1}{2}[0.16_{-0.14}^{+0.16} \pm 0.03] $ & $0.10 \pm 0.04 $ \\
\hline
\hline
\end{tabular}
\end{center}
\end{table}

The unexpected deviation of $f_L$ from the expected value of ${\cal
O}(1)$ is known as the polarization anomaly in $B\to \phi K^*$
decay. In practice, the value of $f_L$ is slightly less than unity
(and predicted to be around 0.9) in the SM. We would like to
mention here that the polarization measurement in all other
vector-vector modes, observed so far, (e.g., $B \to \rho K^*$ and
$B \to \rho \rho $) are in accordance with the SM expectations.

Speculation of the existence of new physics in $B\to \phi K^*$ was
pointed out in \cite{rm1}, in the context of two beyond the
standard model scenarios (namely, R-parity violating supersymmetry
and vector like down quark model). 
The issue of longitudinal polarization problem in $ B \to \phi K^*$
process and its implications were outlined in the review talk
\cite{yg04}.
Recently, there have been a lot
of works in this issue that one can find in the literature 
 \cite{kagan1,co1,hou1,disa1,cheng1,li1,dong1}  both in and beyond
the standard model. To be more specific, what we essentially need
is a destructive longitudinal component, or an enhancement in the
transverse component or the occurrence of both of them coherently
to account for the low longitudinal polarization (or large transverse 
polarization) observed in $B\to
\phi K^*$. But, at this stage, because of our lack of complete
understanding of quark-hadron dynamics, it will be very hard to
pinpoint the exact nature of it. Nevertheless, at least, it will be
immensely rewarding to see if one can really afford to have
similar behaviour in some of the popular beyond the SM scenarios.
Therefore, in this paper we would like to analyze the
decay mode $ \B \to \phi K^* $ in one of the most
popular beyond the SM scenarios, i.e., in the minimal supersymmetric 
standard model with mass insertion approximation \cite{hall86, gabb96}
and  to see whether the observed polarization anomaly 
can be accounted for in this model or not. It should be noted 
here that
the contributions arising from gluonic dipole operator with 
squark-gluino loop, are enhanced by a factor of
$(m_{\tilde g}/m_b)$ compared to the SM contributions due to the
chirality flip from the internal gluino propagator in
the loop and it will interfere destructively with the
longitudinal component of SM amplitude for $RL$ mass mass insertion
(which is what exactly one needs, as noted earlier).
Therefore, one would naively expect that the polarization anomaly in 
$B \to \phi K^*$ mode can possibly be explained by the minimal supersymmetric
standard model with $RL$ mass insertion.

The  paper is outlined as follows. In the next section
we present the basic formalism for the $ B\to V_1 V_2$ decay
mode. In section 3, we calculate the SM contribution in QCD
factorization approach  for the sake of completeness.
Section 4 contains the new physics
contribution to account for the lower longitudinal polarization
and in section 5, we present our conclusions.
\section{Polarization fractions and Triple Product asymmetries in $
B \to V_1 V_2 $ decay}

The most general covariant amplitude for the
decay mode $\Bb \to V_1V_2$ can  be described as \cite{kra}
\be
A(\Bb (p) \to V_1(p_1, \varepsilon_1) V_2(p_2, \varepsilon_2)=
\varepsilon_{1 \mu}^* \varepsilon_{2
\nu}^* \left [ a g^{\mu \nu} + \frac{b}{m_1 m_2} p_1^{\mu}
p_2^{\nu} + \frac{i c}{m_1 m_2} \epsilon^{\mu \nu \alpha \beta} p_{1
\alpha} p_{2 \beta} \right ]\;,\label{bb}
\ee
where $p$ is the $B$ meson momentum and
$m_i$, $p_i$ and $\varepsilon_i$
($i=1,2$) denote the
masses, momenta and polarization vectors of the outgoing vector mesons.

However, it is customary to express the angular distribution of
$\Bb \to V_1 V_2$, with each vector meson subsequently decaying
into two particles, in terms of the helicity amplitudes usually
defined as \be H_{\lambda}= \langle V_1
(\lambda)V_2(\lambda)|{\cal H}_{eff} | \Bb \rangle\;, \ee for
$\lambda=0, \pm 1 $.

The relationship between the helicity amplitudes and the
invariant amplitudes $a$, $b$, and $c$ are given as
\begin{equation}
H_{\pm 1} = a \pm c  \sqrt{x^2 - 1}\;,
~~~
H_0 = - a x - b  (x^2 - 1)\;,
\label{a}
\end{equation}
where $x =(p_1 \cdot p_2)/m_1 m_2 = (m_B^2 - m^2_1 - m^2_2)/(2 m_1 m_2)$.
The corresponding decay rate using the helicity basis 
amplitudes can be given as
\begin{equation}
\Gamma = \frac{p_{c}}{8 \pi m_B^2} \biggr( |H_{+1}|^2 +|H_{-1}|^2+|H_0|^2
\biggr)\;,
\end{equation}
where $p_{c}$ is the magnitude of c.o.m. momentum of the outgoing vector
particles.
It is also convenient to express the relative decay rates
with longitudinal and transverse
polarizations as
\begin{eqnarray}
f_L=\frac{\Gamma_L}{\Gamma} &  = &
\frac{|H_0|^2}{|H_{+1}|^2+|H_{-1}|^2+|H_0|^2}
\;\; , \nonumber \\
f_T=\frac{\Gamma_T}{\Gamma} &  = &
\frac{|H_{+1}|^2+|H_{-1}|^2}{|H_{+1}|^2+|H_{-1}|^2+|H_0|^2}
\;\; .
\end{eqnarray}
The helcity amplitudes ${\bar H}_\l $ for the decay $\B \to {\bar V}_1
{\bar V}_2$, where ${\bar V}_1$ and ${\bar V}_2$ are the antiparticles
of $V_1$ and $V_2$ respectively, have the same decomposition as
(\ref{bb}) with $a \to \bar a$, $b \to \bar b$ and $c \to -\bar c$.
$\bar a$, $\bar b $, and $\bar c$ can be obtained from
$a$, $b$ and $c$ by changing the sign of their weak phases.

To take advantage of more easily extracting the CP odd and CP even
components the angular distribution is often written in the
transversity basis.
The  amplitudes in  transversity  and helicity basis are related to each other
through the  relations
\begin{eqnarray}
A_{\pp} =  \frac{H_{+1} - H_{-1}}{\sqrt{2}},~~~
A_{\pl}  =  \frac{H_{+1} + H_{-1}}{\sqrt{2}},~~~
 A_0  =
 H_0 \label{cb}.
\end{eqnarray}
In the transversity basis the longitudinal and the CP-odd polarizations
are given as
\be
f_L= \frac{|A_0|^2}{|A_\pp|^2+|A_\pl|^2+|A_0|^2}\;,~~~~~~
f_\pp= \frac{|A_\pp|^2}{|A_\pp|^2+|A_\pl|^2+|A_0|^2}\;.
\ee
The triple product asymmetries (TPA's) in $ \Bb \to V_1
V_2$ decays are defined as \cite{datta1}
\bea
{\cal A}_T^{0} &=&
\frac{1}{2}\left [ \frac{{\rm Im}(A_\pp A_0^*)}{\sum_\lambda
|A_\lambda|^{2}}+\frac{{\rm Im}({\bar A}_\pp {\bar A}_0^*)}{\sum_\lambda
|{\bar A}_\lambda|^2}\right ]\;,\nn\\
{\cal A}_T^{\pl} &=& \frac{1}{2}\left [ \frac{{\rm Im}(A_\pp
A_\pl^*)} {\sum_\lambda |A_\lambda|^2}+\frac{{\rm Im}({\bar A}_\pp
{\bar A}_\pl^*)}{\sum_\lambda |{\bar A}_\lambda|^2}\right ]\;, \label{ad}
\eea 
where $\l = 0, \pl, \pp$.

\section{Standard Model contribution}

In the SM, the decay process $\Bb \to \phi \bar K^{*0}$ receives
contribution from the
quark level transition $b \to s \bar s s $, which is induced by the
QCD, electroweak and magnetic penguins. The effective Hamiltonian
describing the decay $b\to s \bar s  s $ 
\cite{beneke1,beneke2} is given as

\be
{\cal H}_{eff}=
\frac{G_F}{\sqrt{2}}V_{qb}V_{qs}^* \biggr[\sum_{j=3}^{10}C_j O_j+C_g O_g
 \biggr],
\ee
where $q=u,~c$.  $O_3, \cdots, O_{6}$ and $O_7,
\cdots, O_{10}$ are the standard model QCD and electroweak penguin operators
respectively, and $O_{g}$ is the gluonic magnetic penguin operator.
The values of the Wilson coefficients at the scale $\mu \approx m_b$
in the NDR scheme are given
in Ref. \cite{buca96} as
\bea
&&C_1=1.082\;,~~~~C_2=-0.185\;,~~~~C_3=0.014\;,~~~~~C_4=-0.035\nn\\
&&C_5=0.009\;,~~~~C_6=-0.041\;,
~~~~C_7=-0.002\alpha\;,~~~~
C_8=0.054 \alpha\;,\nn\\
&&C_9=-1.292\alpha\;,~~~~C_{10}=0.263 \alpha\;,~~~~C_{g}=-0.143\;.
\eea
We use the QCD factorization approach  \cite{beneke1, beneke2}
to evaluate the hadronic matrix elements, which allows us to
compute the nonfactorizable corrections in the heavy quark limit.
Naive factorization is recovered in the heavy quark limit and to
the zeroth order of QCD corrections. The decay mode $B \to \phi K^*$
has been analyzed in Refs. \cite{dong1,das04, cheng2} using the QCD
factorization approach. We will first briefly discuss the
essential differences between these three approaches. It has been 
shown in \cite{cheng2} that the magnetic
dipole penguin will contribute to all the three helicity amplitudes
with almost same order. Later, this result  has been changed in
Ref. \cite{das04} where they have shown that the magnetic
penguin will contribute only to longitudinal polarization amplitude
($H_0$). However, very recently  again it has been 
corrected in Ref.
\cite{dong1} that the positive helicity amplitude $(H_{+1})$ also 
receives a small but nonzero contributions from the
the magnetic dipole operator. It should be noted
here that the contribution of the magnetic 
dipole operator to the tranverse amplitudes are also found to be small
but nonzero in pQCD approach by Li and Mishima \cite{li1}.

Here, we will use the results of QCD factorization method as obtained in 
Ref. \cite{dong1}.   
In this approach the helicity amplitudes are given as  
\bea
H_{0} &=&  -\frac{G_F}{\sqrt 2}~ V_{tb} V_{ts}^*~ \frac{\tilde a^0 f_\phi}
{2 m_{K^*}}
\biggr[(m_B^2 -m_{K^*}^2 -m_\phi^2)(m_B+m_{K^*})A_1^{BK^*}(m_\phi^2)
\nn\\
&& \hspace*{0.2 true cm}
-\frac{4 m_B^2 p_c^2}{m_B+m_{K^*}}A_2^{B K^*}(m_\phi^2) \biggr]\;,\nn\\
H_{\pm 1} &=&- \frac{G_F}{\sqrt 2}~V_{tb} V_{ts}^*~ \tilde a^\pm 
m_\phi f_\phi
\biggr[(m_B+m_{K^*})A_1^{B K^*}(m_\phi^2) \mp \frac{2 m_B p_c}{m_B +
m_{K^*}}V^{B K^*}(m_\phi^2) \biggr]\;,\label{smamp}
\eea
where $\tilde a^h=a_3^h+a_4^h+a_5^h-\frac{1}{2}(a_7^h+a_9^h+a_{10}^h)$
with $h=0, \pm1 $.  $A_{1,2}^{B K^*}(q^2)$ and $V^{BK^*}(q^2)$
are the form factors describing the $B \to K^*$ transitions
\cite{bsw}
evaluated at $q^2=m_\phi^2$.
The expressions for the effective parameters $a_i^h$ appearing 
in the helicity amplitudes
(\ref{smamp}) are given as \cite{dong1}
\bea
a_3^h &=&C_3+\frac{C_4}{N}+\frac{\alpha_s}{4 \pi} \frac{C_F}{N}
C_4 \Big(f_I^h(1)+f_{II}^h(1)\Big) \;,\nn\\
a_4^h &= & C_4+\frac{C_3}{N}+\frac{\alpha_s}{4 \pi} \frac{C_F}{N}
\biggr( C_3 \left [f_I^h(1) +f_{II}^h(1)+ G^h(s_s) + G^h(s_b) \right ]
-C_1\biggr( \frac{\l_u}{\l_t} G^h(s_u)\nn\\
&+ & \frac{\l_c}{\l_t}G^h(s_c)
\biggr) +(C_4+C_6) \sum_{i=u}^b \left (G^h(s_i)
-\frac{2}{3} \right )
+ \frac{3}{2}(C_8+C_{10})\sum_{i=u}^b e_i \left (G^h(s_i)
-\frac{2}{3}\right )\nn\\
&+& \frac{3}{2} C_9 \left [ e_s G^h(s_s)+e_b G^h(s_b)
\right ]
+C_{g} G_{g}^h\biggr)\;,\nn\\
a_5^h &=&C_5+\frac{C_6}{N}-\frac{\alpha_s}{4 \pi} \frac{C_F}{N}
C_6\Big[f_I^h(-1)+ f_{II}^h(-1)\Big]\;, \nn\\
a_7^h &=&C_7+\frac{C_8}{N}-\frac{\alpha_s}{4 \pi} \frac{C_F}{N}
C_8\Big[f_I^h(-1)+ f_{II}^h(-1)\Big] -\frac{\alpha}{9 \pi} N 
C_e^h \;,\nn\\
a_9^h &= & C_9+\frac{C_{10}}{N}+\frac{\alpha_s}{4 \pi} \frac{C_F}{N}
C_{10} \Big[f_I^h(1)+f_{II}^h(1)\Big]- \frac{\alpha}{9 \pi}N C_e^h \;,\nn\\
a_{10}^h &= & C_{10}+\frac{C_9}{N}+\frac{\alpha_s}{4 \pi} \frac{C_F}{N}
C_{9}\Big[f_I^h(1)+f_{II}^h(1)\Big]- \frac{\alpha}{9 \pi} C_e^h
\;,\label{qcd}
\eea
where $\l_q=V_{qb}V_{qs}^*$, $C_F = (N^2 -1)/2N$ and $s_i=m_i^2/m_b^2$.
The QCD penguin loop functions $G^h(s)$ are given by
\bea
G^0(s) &=& \frac{2}{3} -\frac{4}{3} \ln \frac{\mu}{m_b} +4 \int_0^1
dx~ \Phi_\pl^V(x)~ g(x,s)\;,\nn\\
G^{\pm 1 }(s) &=& \frac{2}{3} -\frac{2}{3} \ln \frac{\mu}{m_b} +2 \int_0^1
dx~ \left (g_\pp^{(v) \phi}(x) \pm \frac{1}{4} \frac{dg_\pp^{
(a)\phi}(x)}
{dx} \right )g(x,s)\;, 
\eea 
with the function $g(x,s)$ defined as
\be
g(x,s)=\int_0^1 du~ u(1-u) \ln \left [ s -u(1-u)(1-x) -i \epsilon\right ]\;.
\ee
The EW penguin type
diagrams induced by the operators $O_1$ and $O_2$ are \be C_e^h
=\left ( \frac{\l_u}{\l_t}G^h(s_u)+\frac{\l_c}{\l_t}G^h(s_c)
\right ) \left ( C_2 + \frac{C_1}{N} \right )\;. \ee 
The gluonic dipole
operator $O_g$ gives a tree level contribution as
\bea
G_g^0 &=& -2 \int_0^1 dx~ \frac{\Phi_\pl^\phi(x)}{1-x}\;,\nn\\
G_g^+ &=& \int_0^ 1 dx 
\left (g_\pp^{(v) \phi}(x) + \frac{1}{4} \frac{dg_\pp^{
(a)\phi}(x)}
{dx} \right )
 \frac{1}{1-x}\;,\nn\\
G_g^- &=&0\;.
\eea
The vertex correction factors $f_I^h$
are given as 
\bea
 f_I^0(a)&=& -12 \ln \frac{\mu}{m_b}-18+6(1-a)
+ \int_0^1 dx~ \Phi_\pl^\phi(x)\left ( 3
\frac{1-2x}{1-x}\ln x -
3 i \pi \right ) \;,\nn\\
f_I^{\pm 1 }(a) &=&- 12 \ln \frac{\mu}{m_b}-18+6(1-a)\nn\\
&+&\int_0^1 dx~ \left (g_\pp^{(v) \phi}(x) \pm \frac{a}{4} \frac{dg_\pp^{
(a)\phi}(x)}
{dx} \right )
\left ( 3
\frac{1-2x}{1-x}\ln x - 3 i \pi \right ) \;. 
\eea 
The hard spectator interaction $f_{II}^h$ arising from the
hard spectator interaction with a hard gluon exchange between the vector
meson and the spectator quark of the $B$ meson is given as 
\bea 
f_{II}^0(a) &=&\frac{4 \pi^2}{N}\frac{ i
f_B f_{K^*} f_\phi}{ h_0} \int_0^1 d  \rho \frac{
\Phi_1^B( \rho)}{ \rho} \int_0^1 d v
\frac{\Phi_\pl^{K^*}(v)}{\bar v}
\int_0^1 d u \frac{\Phi_\pl^\phi (u)}{u}\;,\nn\\
f_{II}^{\pm 1}(a) &=& -\frac{4 \pi^2}{N}\frac{2 i  f_B f_{K^*}^\pp 
f_\phi m_\phi}
{ m_B h_{\pm 1}}(1 \mp 1) \int_0^1 d \rho
\frac{\Phi_1^B(\rho)}{
\rho} \int_0^1 d v \frac{\Phi_\pp^{K^*}(v)}{
{\bar v}^2}\nn\\
 &\times & \int_0^1 d u 
\left (g_\pp^{(v) \phi}(u) - \frac{a}{4} \frac{dg_\pp^{
(a)\phi}(u)}
{du} \right )
+ \frac{4 \pi^2}{N} \frac{2 i f_B f_{K^*} f_\phi m_{K^*}
m_\phi}{m_B^2 h_{\pm 1}}
\int_0^1 d  \rho \frac{\Phi_1^B( \rho)}{\rho}\nn\\
&\times & \int_0^1 dvd u 
\left (g_\pp^{(v) {K^*}}(v) \pm \frac{1}{4} \frac{dg_\pp^{
(a){K^*}}(v)}
{dv} \right )\left (g_\pp^{(v) \phi}(u) \pm \frac{a}{4} \frac{dg_\pp^{
(a)\phi}(u)}
{du} \right )
\frac{u+\bar v}{u \bar v^2} \;,
\eea
with $\bar v=1-v $ and 
\bea
h_0 &=& \frac{i f_\phi}{2 m_{K^*}}
\biggr[(m_B^2 -m_{K^*}^2 -m_{\phi}^2)(m_B+m_{K^*})A_1^{B K^*}(m_\phi^2)
-\frac{4 m_B^2 p_c^2}{m_B +m_{K^*}}A_2^{B K^*} (m_\phi^2)\biggr]\;,\nn\\
h_{\pm 1} &=&i f_\phi m_\phi\biggr[(m_B+ m_{K^*})A_1^{B K^*}(m_\phi^2) \mp
\frac{2 m_B p_c}{m_B+m_{K^*}}V^{B K^*}(m_\phi^2)\biggr]\;.
\eea
The asymptotic form of the leading twist 
$(\Phi_\pl^V(x),~\Phi_\pp^V(x))$ and twist-3 
$(g_\pp^{(v)}(x),~g_\pp^{(a)}(x))$
light cone distribution amplitudes
are defined as
\bea
&& \Phi_\pl^V(x)=\Phi_\pp^V(x)=g_\pp^{(a)}(x)=6x(1-x)\;,\nn\\
&&g_\pp^{(v)}(x)= \frac{3}{4}[1+(2x-1)^2]\;.
\eea
The light cone projector for $B$ meson in the heavy quark limit
can be expressed as \cite{beneke1}
\be
{\cal M}^B =-\frac{i  f_B m_B}{4} \left [(1+ \not\!{v}) \gamma_5
 \left \{ \Phi_1^B(\xi)+ {\not\!{n}}_- \Phi_2^B(\xi) \right \} \right
 ]\;,
\ee where $\xi$ is the momentum fraction of the spectator quark in
the $B$ meson, $v=(1,0,0,0)$, $n_-=(1,0,0,-1)$ is the light cone
vector. The normalization conditions are given as \be \int_0^1 d
\xi~ \Phi_1^B(\xi)=1\;,~~~ \int_0^1 d \xi~ \Phi_2^B(\xi)=0\;. \ee
For our numerical evaluation we use \be \int_0^1 d \xi
\frac{\Phi_1^B(\xi)}{\xi}=\frac{m_B}{\lambda_B}\;, \ee with
$\lambda_B=0.46$ GeV, which parametrizes
our ignorance of $B$ meson distribution amplitudes. 

It should be
noted that the presence of logarithmic and linear infrared
divergences in $f_{II}^{\pm 1} $ implies that the spectator
interaction is dominated by the soft gluon exchanges in the final
states. To regulate these divergences, a cutoff parameter of
order $\Lambda_{\rm QCD}/m_b$, with $\Lambda_{\rm QCD}$=0.5 GeV
has been used. 

For numerical analysis, we use the following input parameters. The quark
masses  appearing in the penguin diagrams are pole masses and we have
used the following values (in GeV) as
$m_u = m_d =m_s =0 ,~m_c=1.4$ and $m_b=4.8$.
The decay constants  used are  (in GeV) $f_B=0.161$, $f_{K^*}=0.217$,
$f_\phi=0.231$ and $f_\phi^\pp=0.156$. The form factors are evaluated
in the light-cone sum rule analysis \cite{ball} where the $q^2$ dependence
is given as
\be
F(q^2)= F(0)\exp[c_1(q^2/m_B^2)+c_2(q^2/m_B^2)^2]\;,
\ee
with the parameters as given in Table-2. The particle masses and 
lifetime of $\B $ meson was taken from \cite{pdg}.
For the CKM matrix elements, we have used \cite{pdg}
\begin{eqnarray*}
|V_{cb}|=0.0413 \pm 0.0015\;,~~~~ \bar \rho=0.20 \pm0.09\;,~~~~
\bar \rho = 0.33 \pm 0.05 \;.
\end{eqnarray*}
\begin{table}
\begin{center}
\caption{The parameters of the form factors describing
$B \to K^* $ transitions.}
\vspace*{0.3 true cm}
\begin{tabular}{ccccccc}
\hline
\hline
  && $A_1(0)$&& $A_2(0)$&& $V(0)$\\
\hline
 $F(0)$&& 0.294 &&0.246 && 0.399 \\
 $c_1$&& 0.656 && 1.237 && 1.537 \\
 $c_2$&& 0.456 &&0.822&&1.123 \\
\hline
\hline
\end{tabular}
\end{center}
\end{table}
With these input parameters, we obtain the branching ratio in the
SM as
\be 
{\cal B}(\B \to \phi K^{*0})=(6.34 \pm 0.46) \times 10^{-6}\;, \ee
and the longitudinal and the CP-odd polarizations as \be
f_L=0.89\;,~~~~f_\pp=0.05 \;. \ee 
The triple product asymmetries
${\cal A}_T^{(0,\pl)}$ (\ref{ad}) are found  to be identically zero.

\section{New Physics Contributions}

We now consider the contribution arising from NP. In general the
effective $\Delta B=1$, NP Hamiltonian relevant for the $b \to s
\bar s s$ transition is given as

\be
{\cal H}_{eff}^{NP}\propto \left [\sum_i( C_i^{NP}O_i+
\tilde C_i^{NP} \tilde O_i
) +C_g O_g + {\tilde C}_g {\tilde O}_g \right ] \;,
\ee
where $O_i$ ($O_g$), are the standard model like QCD
(magnetic) penguin operators
with current structure $(\bar s b)_{V-A} (\bar s s)_{V \pm A}$
and $C_i^{NP}$, $C_g^{NP}$ are the
new Wilson coefficients. The operators $\tilde O_i$ $({\tilde O}_g)$
are obtained from $O_i$ $(O_g)$  by exchanging $L \leftrightarrow
R$. As discussed in Ref. \cite{kagan1} the NP contributions to the different
helicity amplitudes are given as
\bea
A^{NP}(\Bb \to \phi K^{*0})_{0,\pl} &\propto & C_i^{NP} - {\tilde C}_i^{NP}\;,
\nn\\
A^{NP}(\Bb \to \phi K^{*0})_{\pp} & \propto & C_i^{NP} + {\tilde C}_i^{NP}\;.
\eea
Thus, in the presence of new physics, the different amplitudes
can be given as
\bea
A_{0, \pl}&=&A_{0,\pl}^{SM}+A_{0,\pl}^{NP}
=A_{0,\pl}^{SM}\left [1+ e^{i \phi_N}
(r_{0,\pl} - {\tilde r}_{0,\pl})\right ]\;,\nn\\
A_\pp&=& A_\pp^{SM}+A_\pp^{NP}=A_\pp^{SM}\left [1+ e^{i \phi_N}
(r_\pp + {\tilde r}_\pp) \right ]\;, \label{cs} \eea 
where
$r_{\l}$, with $(\l=0,\pl,\pp)$ are the ratio of NP (arising from $C_i O_i$
and $C_g O_g$ part of the Hamiltonian) to SM amplitudes, 
${\tilde r}_\l$ are the
corresponding values arising from the $\tilde C_i \tilde O_i $ 
and ${\tilde C}_g {\tilde O}_g$ part.
 $\phi_N $ is the relative weak phase between the SM and NP
amplitudes.
For simplicity, we have assumed a common weak phase for the $C$ and
${\tilde C}$ contributions and zero strong phase between the SM
and the NP amplitudes.

Thus the branching ratio is given as 
\be 
{\cal B}(\B \to \phi
K^{*0})={\cal B}^{SM} \left [ 1+ \frac{\sum_\l R_\l^2 |A_\l^{SM}|^2}{
\sum_\l  |A_\l^{SM}|^2}+ 2 \frac{\sum_\l R_\l |A_\l^{SM}|^2}{
\sum_\l  |A_\l^{SM}|^2} \cos \phi_N \right ]\label{br}\;, \ee
where $R_{\pl,0}=r_{\pl,0}-{\tilde r}_{\pl,0}$ and $R_\pp=
r_\pp+\tilde r_\pp$ and ${\cal B}^{SM}$ denotes the SM branching ratio.
 The longitudinal and the CP-odd polarizations
now read as 
\bea 
f_L &= & \frac{|A_0^{SM}|^2 \left [1+R_0^2+2 R_0 \cos
\phi_N \right ]}{
\sum_\l |A_\l^{SM}|^2 \left [1+R_\l^2+2 R_\l \cos \phi_N \right ]}\;,\nn\\
f_\pp &= & \frac{|A_\pp^{SM}|^2\left [1+R_\pp^2+2 R_\pp \cos \phi_N
\right ]}{
\sum_\l|A_\l^{SM}|^2\left [1+R_\l^2+2 R_\l \cos \phi_N
\right ]}\;.
\label{pol}
\eea
Furthermore, in the presence of NP,
the Triple Product asymmetries (\ref{ad}) are given as
\bea
{\cal A}_T^{0} &= &\frac{2(R_\pp-R_0)\sin \phi_N}{
\sum_\l|A_\l^{SM}|^2\left [1+R_\l^2+2 R_\l \cos \phi_N 
\right ]}\;,\nn\\
{\cal A}_T^{\pl} &= &\frac{2(R_\pp-R_\pl)\sin \phi_N}{
\sum_\l|A_\l^{SM}|^2\left [1+R_\l^2+2 R_\l \cos \phi_N
\right ]}\;.
\label{tpa}
\eea

We now analyze the decay process $\B \to \phi K^{*0} $ in the minimal
supersymmetric standard model (MSSM) with mass insertion approximation.
This decay mode receives supersymmetric (SUSY)
contributions mainly from penguin and box diagrams containing
gluino-squark, chargino-squark and
charged Higgs-top loops. Here, we consider only the
gluino contributions, because
the chargino and charged Higgs loops are expected to be suppressed
by the small electroweak gauge couplings.
However, the gluino mediated FCNC contributions are
of the order of strong interaction strength, which may exceed the
existing limits.  Therefore, it is customary to rotate the
effects, so that the FCNC effects occur
in the squark propagators rather than in
the couplings and to parameterize them
in terms of dimensionless parameters. Here we work in the usual mass insertion
approximation \cite{hall86, gabb96}, where the flavor mixing
$i \to j$ in the down-type squarks associated with $\tilde q_B$ and
$\tilde q_A$ are parametrized by  $(\delta^d_{AB})_{ij}$, with
$A,~B=L,~R$ and $i,j$ as the generation indices. More explicitly
$(\delta^d_{LL})_{ij}
=({V_L^d}^\dagger M_{\tilde d}^2 V_L^d)_{ij}/ m_{\tilde q}^2$, where
$M_{\tilde d}^2$
is the squared down squark mass matrix and $m_{\tilde q}$ is
the average squark mass.
$V_d$ is the matrix which diagonalizes the down-type quark mass matrix.

Thus, the new effective $\Delta B=1$ Hamiltonian relevant for the $\B \to
\phi K^*$
process arising from new penguin/box diagrams with gluino-squark in the
loops is given as
\be
{\cal H}_{eff}^{SUSY} = -\frac{G_F}{\sqrt 2} V_{tb}V_{ts}^*
\left [\sum_{i=3}^6 \left ( C_i^{NP}O_i+ \tilde C_i^{NP} \tilde O_i
\right )
+C_g^{NP} O_g + \tilde C_g^{NP} \tilde O_g \right ]\;,
\ee
where $O_i$ $(O_g)$ are the QCD (magnetic) penguin operators and
the $C_i^{NP}$ ($C_g$) are the
new Wilson coefficients. The operators $\tilde O_i$
are obtained from $O_i$  by exchanging $L \leftrightarrow
R$.

To evaluate the amplitude in the MSSM, we have to first determine
the Wilson coefficients at the $b$ quark mass scale. At the
leading order, in mass insertion approximation, the new Wilson
coefficients corresponding to each of the operator at the scale
$\mu \sim \tilde m \sim M_W$ are given as \cite{gabb96,ko04} \bea
C_3^{NP} & \simeq & -\frac{\sqrt 2 \alpha_s^2}{4 G_F V_{tb}
V_{ts}^* m_{\tilde q}^2}\left ( \delta_{LL}^d \right )_{23} \left
[ - \frac{1}{9}B_1(x) -\frac{5}{9} B_2(x)-\frac{1}{18}P_1(x)
-\frac{1}{2}P_2(x) \right ]\;,\nn\\
C_4^{NP} & \simeq & -\frac{\sqrt 2 \alpha_s^2}{4 G_F V_{tb}
V_{ts}^* m_{\tilde q}^2}\left ( \delta_{LL}^d \right )_{23}
\left [ - \frac{7}{3}B_1(x) +\frac{1}{3} B_2(x)+\frac{1}{6}P_1(x)
+\frac{3}{2}P_2(x) \right ]\;,\nn\\
C_5^{NP} & \simeq & -\frac{\sqrt 2 \alpha_s^2}{4 G_F V_{tb}
V_{ts}^* m_{\tilde q}^2}\left ( \delta_{LL}^d \right )_{23}
\left [  \frac{10}{9}B_1(x) +\frac{1}{18} B_2(x)-\frac{1}{18}P_1(x)
-\frac{1}{2}P_2(x) \right ]\;,\nn\\
C_6^{NP} & \simeq & -\frac{\sqrt 2 \alpha_s^2}{4 G_F V_{tb}
V_{ts}^* m_{\tilde q}^2}\left ( \delta_{LL}^d \right )_{23}
\left [ - \frac{2}{3}B_1(x) +\frac{7}{6} B_2(x)+\frac{1}{6}P_1(x)
+\frac{3}{2}P_2(x) \right ]\nn\\
C_{g}^{NP} &\simeq &
-\frac{2\sqrt 2 \pi \alpha_s}{2 G_F V_{tb}
V_{ts}^* m_{\tilde q}^2}\biggr[ \left ( \delta_{LL}^d \right )_{23}
\left (  \frac{3}{2}M_3(x) -\frac{1}{6} M_4(x)\right )\nn\\
&&\hspace*{1.0 true in}
+ \left ( \delta_{LR}^d \right )_{23}
\left (\frac{m_{\tilde g}}{m_b}\right )
\frac{1}{6}\left (4B_1(x)-\frac{9}{x}B_2(x) \right )\biggr]\;,
\eea
where $x=m_{\tilde g}^2/m_{\tilde q}^2 $. The loop functions 
appearing in these
expressions can be found in Ref. \cite{gabb96}.
The corresponding $ \tilde C_i^{NP}$ are obtained from $C_i^{NP}$ by
interchanging $L \leftrightarrow R$.
It should be noted that the $(\delta_{LR}^d)_{23}$ contribution
is enhanced by $(m_{\tilde g}/m_b)$ compared to that of 
the SM and the $LL$ insertion due to the chirality flip from the
internal gluino propagator in the loop. Therefore, the magnetic dipole
operators in supersymmetric model are found to contribute
significantly.

The Wilson coefficients at low  energy
$C_i^{NP}(\mu \sim m_b)$,  can be obtained from $C_i^{NP}(M_W)$
by using the Renormalization Group (RG) equation, as discussed in
Ref. \cite{buca96}, as
\be
{\bf C}(\mu) ={\bf U}_5(\mu, M_W) {\bf C}(M_W)\;,
\ee
where ${\bf C}$ is the $6 \times 1$ column vector of the
Wilson coefficients and
${\bf U}_5(\mu, M_W)$ is the five-flavor $6 \times 6$ evolution matrix.
In the next-to-leading order (NLO), ${\bf U}_5(\mu, M_W)$ is given by
\be
{\bf U}_5(\mu, M_W)=\left (1+\frac{\alpha_s(\mu)}{4 \pi} {\bf J} \right )
{\bf U}_5^{(0)}(\mu, M_W)\left (1-\frac{\alpha_s(M_W)}{4 \pi} {\bf J}
\right )\;,
\ee
where ${\bf U}_5^{(0)}(\mu, M_W)$ is the leading order (LO)
evolution matrix and ${\bf J}$ denotes the NLO corrections to the evolution.
The explicit forms of ${\bf U}_5(\mu, M_W)$ and
${\bf J}$ are given in Ref. \cite{buca96}.

Since the $O_g$ contribution to the matrix element is $\alpha_s$
order suppressed, we consider only leading order RG  effects for
the coefficient $C_g^{NP}$, which is given as \cite{ko04} 
 \be
C_g^{NP}(m_b)\simeq 
-0.15+0.70~ C_{g}^{NP}(M_W)\;. \ee
For the numerical analysis, we fix the SUSY parameter as
$m_{\tilde q} =m_{\tilde g}=500 $ GeV, $\alpha_s(M_W)=0.119$.
The absolute values of the mass insertion parameters
$(\delta_{AB}^d)_{23}$, with $A,B=(L,R)$ are constrained by the
experimental value of $B \to X_s \gamma $ decay \cite{gabb96}. 
These constraints are very weak for $LL$ and $RR$ mass
insertions and the existing limits come only from their definitions
$|(\delta_{LL,RR}^d)_{23}|<1$. The $LR$ and $RL$ mass insertions
are more constrained and for instance with $m_{\tilde g}\simeq
m_{\tilde q} \simeq 500$GeV, we have $|(\delta_{LR,RL}^d)_{23}|
\leq 1.6 \times 10^{-2}$. In our analysis, we will use the above
bounds on the mass insertion parameters, i.e., 
\be
|(\delta_{LL,RR}^d)_{23}| <1~~~~{\rm
and}~~~~|(\delta_{LR,RL}^d)_{23}| \leq 1.6 \times 10^{-2}\;. \ee
Now substituting the values of the RG evoluted Wilson coefficients
$C_i^{NP}(m_b)$'s  in Eq. (\ref{qcd}) we obtain the corresponding
$a_i^h$'s and hence with Eqs. (\ref{cb}), (\ref{smamp}) 
and (\ref{cs}) the amplitudes. Assuming
that all the mass insertion parameters $ (\delta^d_{AB})_{23}$
have a common weak phase, we obtain the new physics
 parameters arising from the $LL(LR)$ and $RR(RL)$ mass
insertions as 
\bea 
&&(r_{0})_{LL}= ({\tilde r}_0)_{RR}
<0.44\;,~~~~~~~~~~~~~~~~(r_{0})_{LR}= ({\tilde r}_0)_{RL} \leq 1.3\;,
\nn\\
&& (r_{\pl})_{LL}
= ({\tilde r}_\pl)_{RR} < 6.2 \times 10^{-2}\;,
~~~~~~~~(r_{\pl})_{LR}= ({\tilde r}_{\pl})_{RL} \leq 7.0 \times
10^{-2}\;,\nn\\
&&(r_{\pp})_{LL}= ({\tilde r}_\pp)_{RR} < 6.0 \times 10^{-2}\;,
~~~~~~~(r_{\pp})_{LR}= ({\tilde r}_\pp)_{RL} \leq
7.2 \times 10^{-2}\;. 
\label{po}\eea 
Let us now analyze the variation of CP-odd polarization fraction and 
branching ratio in the presence of new physics. We first consider the 
contributions arising from the $LL$ and $RR$ mass insertions. As seen from Eq. 
(\ref{po}), these contributions are quite small and it is expected that
they cannot accommodate the observed large CP-odd polarization.
Now using the maximum  values of $r_\lambda$ from (\ref{po}), we
plot the CP odd polarization ($f_\pp$) (\ref{pol}) and the 
branching ratio (\ref{br})
versus the weak phase $\phi_N $ for three different cases ($LL$, $RR$
and in the presence of both $LL$ and $RR$ contributions). It is seen from 
Figure-1 that, indeed the large
CP odd polarization fraction cannot be accommodated in these cases although
the observed branching ratio can be accommodated with $LL$ or $RR$ mass
insertions. Next we consider the contributions arising from 
$LR$, $RL$ and the simultaneous presence of $LR$ and $RL$ mass insertions. 
As seen from Figure-2, in this case  
the observed CP-odd polarization fraction ($f_\pp$) and 
the longitudinal polarization fraction ($f_L$) can be accommodated 
with either $LR$ or $RL$ mass insertions. The branching ratio can 
also be accommodated 
with these mass insertions as seen from Figure-3.
\begin{figure}[htb]
   \centerline{\epsfysize 3.25 truein \epsfbox{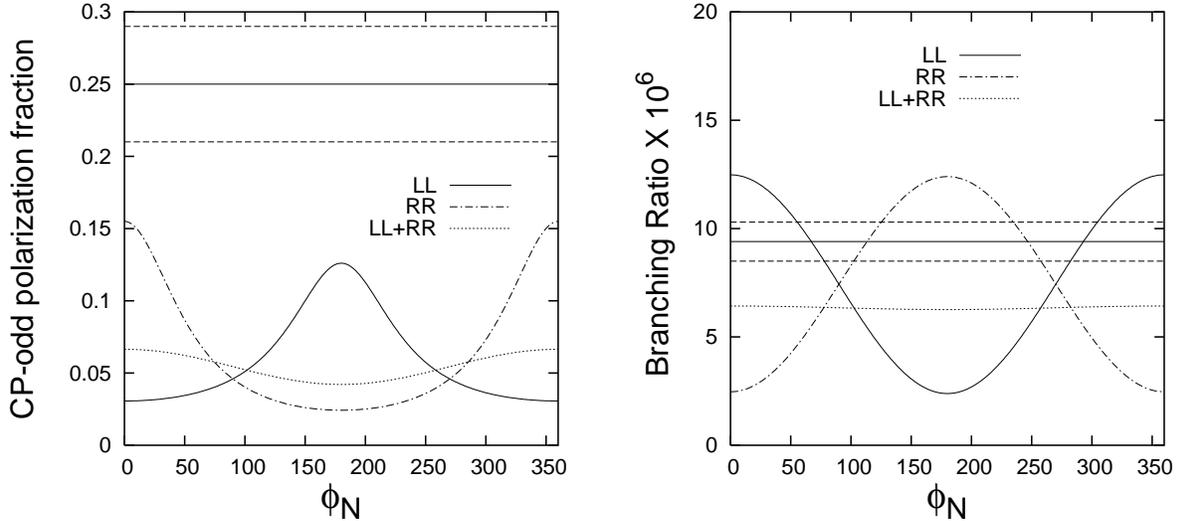}}
 \caption{
  The CP-odd polarization fraction ($f_\pp$) and the branching ratio
of  the process $\B \to \phi K^{*0}$ with $LL$ and $RR$ mass insertions
 versus the weak phase
 $\phi_{N}$ (in degree). The horizontal solid lines represent the
experimental central value and the dashed lines represent the
$1 \sigma $ range.}
  \end{figure}
\begin{figure}[htb]
   \centerline{\epsfysize 3.25 truein \epsfbox{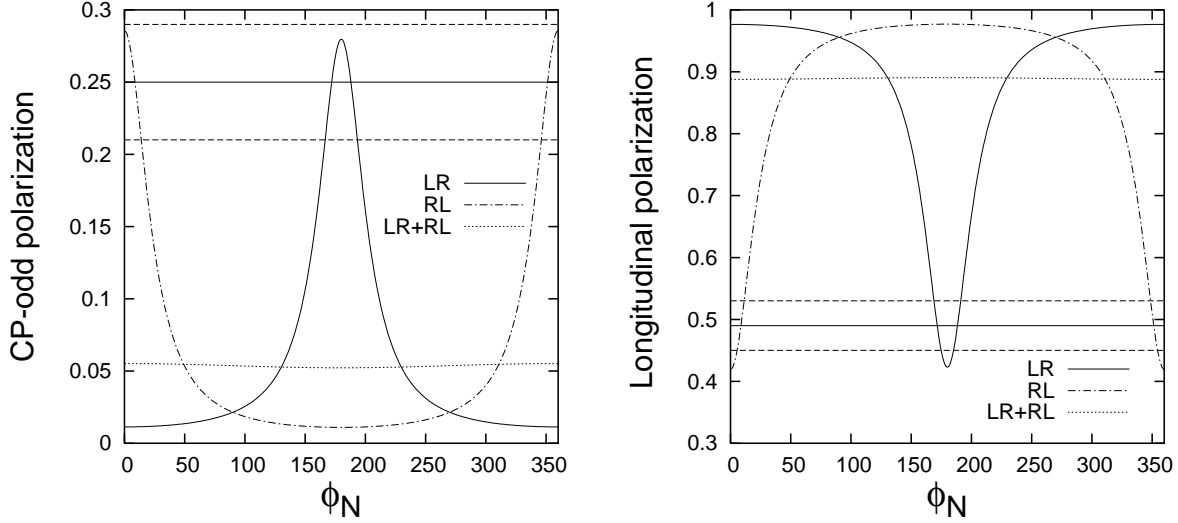}}
 \caption{The CP-odd and the logitudinal  polarization fractions
of  the process $\B \to \phi K^{*0}$
 with $LR$ and $RL$ mass insertions.}
  \end{figure}
\begin{figure}[htb]
   \centerline{\epsfysize 2.8 truein \epsfbox{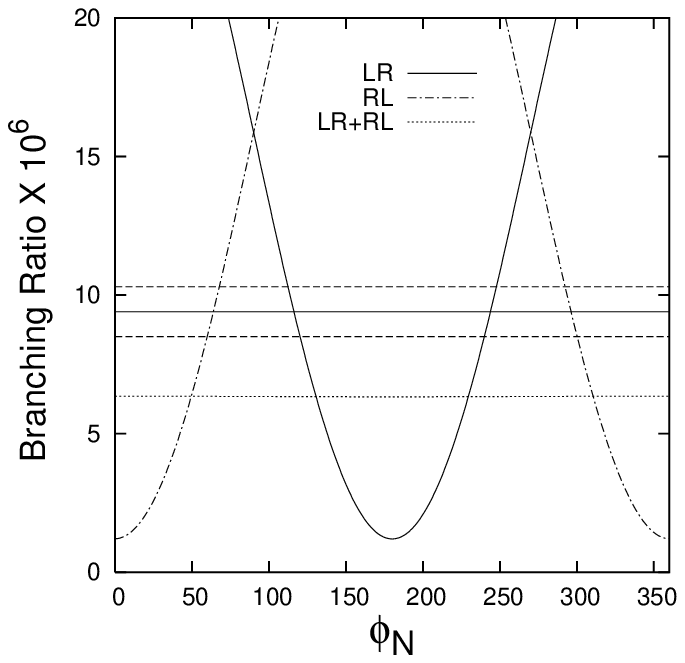}}
 \caption{The branching ratio of  the process $\B \to \phi K^{*0}$
 with $LR$ and $RL$ mass insertions.}
  \end{figure}

\section{Conclusions}
Observation of unexpectedly smaller longitudinal polarization 
(and large transverse polarization) in
the penguin dominated $B \to \phi K^*$ mode
poses a serious challenge both to the theorists and
experimentalists in B-physics. This has in turn ignited the desire
of revealing the existence of new physics beyond the SM. While at
present there is no clear indication of any NP but with the
accumulation of more data, if the longitudinal polarization
stabilizes around the present central value, i.e., $f_L = 0.5$,
with {\it reduced error bars}, then this might be the
first clear evidence of new physics in the $b \to s$
penguin decay amplitudes. 
The polarization measurements in various
vector-vector modes undertaken by the ongoing B-factory
experiments and the experiments to be performed at BTeV and LHC-b
will definitely be able to guide us to understand the dynamics and
possibly provide us a meaningful answer to our quest for the
existence of new physics.

We have employed supersymmetry with mass insertion approximation
and shown that the low longitudinal polarization in $\B \to 
\phi K^{*0}$ can be
accommodated in this beyond the standard model scenario with 
either $LR$ or $RL$ mass insertion approximation. 

Since in the B-factory data so far we have only seen some kind of
deviation in the $b\to s \bar s s$ transition measurements it may
be worthwhile to continue our effort in this direction and check
carefully if we can really observe NP in this type of penguin
induced transitions. If NP is present in the $b\to s\bar s s$
transitions and indeed if it is responsible for the observed lower
longitudinal polarization then we expect to see the same
effect of lower longitudinal polarization in another charmless
vector-vector mode, i.e., $B_s \to \phi\phi$, which is governed by the
same penguin induced $b\to s \bar s s$ transition. Already the
branching ratio for this mode  has been measured by the CDF
Collaboration \cite{res9,cdf05} and  we are looking forward to
the polarization measurements in this mode.
This in turn, at
least, will provide us a clear picture of the charmless vector-vector
transitions induced by the $b\to s\bar s s$ penguins and possibly
revealing the existence of NP in
penguin induced $b\to s\bar s s$ transitions. If confirmed, the
polarization anomaly along with the deviation measured in ${\rm
sin} 2\beta$ measurements may bring us one step further towards
the establishment of NP in the penguin induced $b\to s\bar s  s$
transitions. It is therefore urgently needed to closely examine
experimentally all the possible charmless vector-vector modes to
confirm or rule out the existence of new physics.

{\bf Acknowledgments}

AKG was supported by the National Science Council of the Republic
of China under Contract Number NSC-93-2811-M-007-060. RM was
partly supported by the Department of Science and Technology,
Government of India, through Grant No. SR/FTP/PS-50/2001.


\end{document}